\documentclass[aps,prb,superscriptaddress,letterpaper,twocolumn,showpacs,amsmath,amssymb,byrevtex]{revtex4}

\usepackage[latin1]{inputenc}
\usepackage{graphicx}
\usepackage{color}
\usepackage{dcolumn}
\usepackage{bm}
\usepackage{hyperref}
\usepackage{times}

\begin{document}
\title{Exact Kohn-Sham eigenstates versus quasi-particles in simple models of strongly correlated electrons}

\author{D. J. Carrascal}
\affiliation{Departamento de F\'{\i}sica, Universidad de Oviedo, 33007 Oviedo, Spain}
\affiliation{Nanomaterials and Nanotechnology Research Center, CSIC - Universidad de Oviedo, Spain}

\author{J. Ferrer}
\affiliation{Departamento de F\'{\i}sica, Universidad de Oviedo, 33007 Oviedo, Spain}
\affiliation{Nanomaterials and Nanotechnology Research Center, CSIC - Universidad de Oviedo, Spain}
\affiliation{Department of Physics, Lancaster University, UK}

\date{\today}

\begin{abstract}
We present analytic expressions for the exact density functional and Kohn-Sham Hamiltonian of simple
tight-binding models of correlated electrons. These are the single- and double-site versions of the
Anderson, Hubbard and spinless fermion models. The exact exchange and correlation potentials keep the full
non-local dependence on electron occupations. The analytic expressions allow to compare the Kohn-Sham eigenstates of exact density functional
theory with the many-body quasi-particle states of these correlated-electron systems. The exact Kohn-Sham
spectrum describes correctly many of the non-trivial features of the many-body quasi-particle spectrum,
as for example the precursors of the Kondo peak. However, we find that some pieces of the quasi-particle
spectrum are missing because the many-body phase-space for electron and hole excitations is richer.
\end{abstract}

\pacs{71.15.Mb, 71.10.Fd}

\maketitle

\section{Introduction}

Density Functional Theory \cite{Hohenberg64,Kohn65} allows to tackle
complex quantum systems comprising $N$ interacting electrons. Its
essence consists on the replacement of the extremely convoluted many-particle
electronic interactions with an effective one-body potential, also
known as the exchange and correlation potential $V^{XC}$, whereby
the $N$-particle Hamiltonian is substituted by a fictitious one-particle
Kohn-Sham Hamiltonian $H^{KS}$. The exact $V^{XC}$ is however not
known and it is a widespread belief that it is not possible to find
an analytic exact expression for it. The popularity of Density Functional
Theory (DFT) has arisen from the fact that semi-empirical fittings
of $V^{XC}$ to the exchange and correlation potential of the Jellium
model in the Local Density Approximation (LDA) and improvements over
it\cite{Perdew81,Perdew96,Becke93} perform remarkably well for a
large majority of materials, molecules and nanostructures. The qualitative
features of the structural and functional properties of many systems
are usually well reproduced, and in a number of cases, quantitative
agreement can also be reached. However, these practical implementations
of DFT are not perfect. They fail to predict a number of relevant
properties, specially for strongly correlated electronic systems.
Yang and coworkers have discussed some explicit conditions that exact
energy functionals must obey\cite{Cohen08b,Mori09}.

DFT has also been proposed for tight-binding models of strongly correlated
electrons\cite{Schonhammer95}. The availability of exact semi-analytical
or numerical results for the ground state energy as a function of
the electron concentration in the Hubbard and the spinless fermion
models\cite{Lieb68,Yang66} has allowed to establish a Bethe ansatz
LDA theory for them\cite{Schonhammer95,Lima03,Schenk08}. An extension
of the theory to describe time-dependent external potentials has enabled
the description of non-equilibrium electron transport phenomena\cite{Verdozzi08,Kurth10,Burke11a}.
However, Bethe ansatz LDA theory also has limitations. First, since
the Bethe Ansatz solution expresses the ground-state energy in terms
of the electron concentration, only a local density approximation
could be formulated. Second, the analytic formula for ground state
energy is exact only at half-filling, while away from it semi-analytical
or numerical fittings to the solution of the Bethe Ansatz equations
must be performed. Finally, inhomogeneous systems where strong correlations
take place in localized region of them are better described by Anderson
models. However, the Anderson model is Bethe Ansatz-solvable only
if the band of uncorrelated electrons is linearized\cite{Tsvelick83},
leaving the system energy unbounded from below. Therefore, the ground
state energy can not be obtained by a minimization procedure, which
renders the Bethe ansatz LDA approach useless for the Anderson model.

The quasi-particle (QP) excitation spectrum of a system determines
its response to external perturbations according to Landau's Fermi
liquid theory. Furthermore, this spectrum is directly accessible via
spectroscopic techniques of different sorts. It would therefore be
quite useful if the Kohn-Sham (KS) eigenstates provided at least
a qualitative description of it, as one would expect to happen at
least for systems where electronic correlations are weak. This is
indeed confirmed by a vast amount of calculations and comparisons
between KS eigenvalues and experimental or numerical data of weakly
correlated materials. However, quantitative agreement is sometimes
not so good. Furthermore the KS spectrum is frequently qualitatively
wrong in strongly-correlated materials. Notice now that even if the
exact $V^{XC}$ of a specific system is known, a possible correspondence
between the exact KS and the exact many-body QP
spectra is not supported at all by the basic theorems of DFT.
An exception is the Highest Occupied Molecular Orbital (HOMO) which
by Janak's theorem is given by the chemical potential of the system,
which is a ground-state property\cite{Janak78,Perdew83,Sham83}. In
other words, DFT predicts the correct position of the HOMO level of
a system, provided that the exact $V^{XC}$, or a very good approximation
to it, is known. Failures to predict the correct position of the HOMO
must therefore be attributed to a poor approximation to the exact
$V^{XC}$. Failures to reproduce the rest of the spectrum could however
be due either to limitations of DFT proper, or to a poorly approximated
functional. Indeed, while an exact functional may not provide a good
description of the full QP spectrum, it is clear that if in addition,
the quality of the approximate functional is poor, the proposed spectrum
of KS eigenvalues will bear a small resemblance to the true QP spectrum.
Since no exact functional for a strongly correlated system has ever been
developed, the above two sources of disagreement have never been fully
disentangled. The main goal of this article is to separate them.
We will find the exact KS eigenvalues of several simple models of
strongly correlated electrons and compare them with the exact
many-body QP spectrum. This will allow us to understand the size of
the self-energy corrections to the exchange-correlation potential.

One of the main sources of disagreement between approximate DFT KS
eigenvalues and exact QP originates in the mean-field-like treatment of electronic
correlations which lie at the heart of LDA. Indeed, electrons behave
as quantum point particles. However, mean-field theories replace quantum
probabilities by classical clouds of charge. As a result, every electron
may interact with its own charge cloud, giving rise to spurious direct
and exchange self-interaction effects. Additionally, each electron
interacts with the clouds of other electrons having opposite spin,
leading to what is sometimes called the static correlation error\cite{Cohen08b}.
For systems containing more than one atom, these mean field clouds
are spread throughout the whole entity in contrast to electrons which
are always point particles and therefore fully localized. This spread
gives rise to further spurious effects termed delocalization
errors, which lead to incorrect dissociation energies and QP excitation
energies for molecules\cite{Cohen08b}.
A prototypical example of the delocalization error is an $H_2^+$ molecule
in the dissocation limit where the two ions are held widely apart.
The single electron in the molecule has equal probability of residing in
any of the two atoms, but a measuring proccess will find it fully localized in
only one of them. Mean field theories in contrast place half an electron
in each ion. The excitation energy of an added quasi-electron will therefore
be different in the two cases\cite{Romaniello09,Explanation}.

Improving the description of the QP spectrum therefore implies improving
the description of electronic correlations. The Hartree-Fock approximation
as well as the self-interactions correction scheme\cite{Perdew81,Rieger95,Filippetti03,Toher09}
get rid of the self-interaction effect, but not of other mean-field
drawbacks. To go beyond these schemes, the Dyson-Sham-Schl\"uter equation
must be used
\begin{equation}
G=G_{approx}^{KS}+G_{approx}^{KS}\,(\Sigma^{XC}-V_{approx}^{XC})\, G
\end{equation}
where $G_{approx}^{KS}$ is the Green's function obtained from the
approximate KS Hamiltonian. Notice that $G_{approx}^{KS}$ carries
already a mean-field description of the electron interaction. A perturbative
expansion for the self-energy $\Sigma^{XC}$
must then be set up to improve the description of correlations and in particular
to amend the destruction of quantum effects brought about by the mean-field
approximation. The GW approximation\cite{Hedin65} has been quite
successful in the description of electronic and optical properties
directly linked to the QP spectrum\cite{Aulbur99,Onida02}, but does
not correct the problems mentioned above. Some recent work by Romaniello
and coworkers show how the careful inclusion of vertex corrections
allows to get rid not only of the self-interaction effects but also
of part of the delocalization effects\cite{Onida02,Bruneval05,Romaniello09}.
However, delocalization debris remains since molecular dissociation
is still not well handled.
In addition, Millis and coworkers\cite{Wang08,Thygesen08} have studied
the performance of the GW approximation for the Anderson model, and
shown how this approximation can describe Coulomb blockade effects,
but fails to describe the emergence of Kondo Physics\cite{HewsonBook,Ferrer87}.
Dynamical Mean Field Theory, implemented together with an accurate
impurity-solver, includes many of the most relevant short-range correlation
effects\cite{Sun04,Kotliar06,Jacob10,Korytar11}.

We have devised a procedure that has allowed us to find analytic expressions
for the exact energy density functional of the single- and double-site
Anderson, Hubbard and spinless fermion models, from which we have
been able to write down the corresponding exact Hamiltonians $H^{KS}$.
Since the QP spectrum of these models is available analytically
from conventional many-body techniques, we have been able to perform
explicit and detailed comparisons of the full spectra of exact
KS eigenvalues and of exact many-body QP. We have found that the
KS eigenvalue corresponding to the HOMO level agrees with the
corresponding QP state. This implies that the exact $H^{KS}$ correctly
predicts that the lowest energy for electron addition of an $N$-electron system
is equal to the highest energy for electron removal of the corresponding $N+1$
system\cite{Romaniello09}. We have also found that the exact $H^{KS}$ of
the Anderson model describes correctly the emergence of the Kondo resonance and of
other quasi-particles. However, we find that there exact density functional theory
misses pieces of the exact many-body QP spectrum. A way to improve
the description of the spectrum would be to use again the Dyson-Sham-Schl\"uter
equation
\begin{equation}
G=G_{exact}^{KS}+G_{exact}^{KS}\,(\Sigma^{XC}-V_{exact}^{XC})\, G
\end{equation}
where $G_{exact}^{KS}$ is the Green's function associated to the
exact KS Hamiltonian. We expect that this self-energy and its perturbative
expansion should be much simpler than the self-energy defined in Eq.(1) above,
because now the unperturbed Green's function retains the full quantum
nature of electrons.
Our piece of work is complementary to efforts by other groups to provide
exact functionals for simplified systems. We mention here recent work
by Burke and collaborators, who have found numerically exact density
functionals for some one-dimensional models by combining DFT with
Density Matrix Renormalization Group techniques\cite{Burke11b}.

The layout of this article is as follows. Section II describes the
methodology employed to find out exact functionals for systems with
a small number of electrons. This methodology is applied in sections
III and IV to describe the single-site Hubbard model, and the double-site
Anderson model, respectively. The conclusions are laid down in section
V. The solution of the double-site Hubbard model is placed in appendix
A. The solution of the double-site spinless fermion model can be found
in appendix C.

\section{Methodology}

We begin with a description of our method, which is based on the formulation
of DFT on a lattice\cite{Schonhammer95}. We have found that the conventional
ensemble-based method to describe non-integer occupations\cite{Levy79,Perdew83}
fails in the formulation of the exact density functional of the single-site
model described below. We have therefore devised an alternative method
which is specifically adapted for the description of quantum systems
with a small but not necessarily integer number of electrons $N$.

We consider a physical system whose time-evolution is dictated by
a tight-binding Hamiltonian. As an example, we write explicitly the
Hamiltonian of the Anderson model,
\begin{eqnarray}
\hat{H}= &  & \sum_{i,\sigma}\,\epsilon_{c}\,\hat{n}_{c,i,\sigma}\,+\,\sum_{\sigma}\,\epsilon_{d}\,\hat{n}_{d,\sigma}\,-\,\sum_{i,\sigma}\, t_{0}\,(\hat{c}_{i,\sigma}^{\dagger}\,\hat{c}_{i+1,\sigma}+h.c.)\,\nonumber \\
 &  & -\, t\,\sum_{\sigma}\,(\hat{c}_{1,\sigma}^{\dagger}\,\hat{d}_{\sigma}+h.c.)\,+\, U\,\hat{n}_{d,\uparrow}\,\hat{n}_{d,\downarrow}
\end{eqnarray}
 where a set of $N$ electrons hop back and forth along a chain of
$i=1,..,{\cal M}$ atoms, labeled by the index $c$, and to another
atom, denoted by the index $d$, where electron correlations take
place via a Coulomb term $U$. The $\sigma-$index denotes the up
and down components of the electron spin.

We use the Fock space of states of the system $\{|\,\phi>\}$ to set
up our variational scheme. Site occupations, electron numbers and
the expectation value of the Hamiltonian are given by
\begin{eqnarray}
n_{\alpha,\sigma}(\phi)\, & = & \,\frac{<\phi\,|\,\hat{n}_{\alpha,\sigma}\,|\,\phi>}{<\phi\,|\,\phi>}\nonumber \\
N_{\sigma} & = & \sum_{i}n_{c,i,\sigma}+n_{d,\sigma}\nonumber \\
E(\phi)\, & = & \,\frac{<\phi\,|\,\hat{H}\,|\,\phi>}{<\phi\,|\,\phi>}
\end{eqnarray}
We wish to define an energy density functional $Q[n_{c,i,\sigma},n_{d,\sigma'},U]$
whose minimization gives the exact ground state energy $E^{0}$ and
occupations $n_{i,\sigma}^{0}$ for a target set of electron numbers
$(N_{\uparrow}^{0},N_{\downarrow}^{0})$. To define $Q$, we note
that every given set of occupations $\{n_{c,i,\sigma},n_{d,\sigma'}\}$
can be reproduced by several states $|\,\phi>$. In other words, if
we classify these states in boxes labeled by each occupation set,
then each box contains several $|\,\phi>$, and each of these has
a different energy $E(\phi)$. However, if we choose in each box the state
$|\,\phi^{m}>$ with minimum energy $E^{m}=E(\phi^{m})$, we achieve
a one-to-one correspondence between occupation sets and energies for every box,
which allows to define the energy density functional
$Q[n_{c,i,\sigma},n_{d,\sigma'},U]=E^{m}$\cite{Provided}. Since there exist
in general several sets of occupation numbers $\{n_{c,i,\sigma},n_{d,\sigma'}\}$
giving the same target electron numbers $N_{\sigma}=N_{\sigma}^{0}$,
the ground state energy $E^{0}$ is obtained by minimizing $Q$ over
all those sets. This procedure then defines the ground state occupations
$\{n_{c,i,\sigma}^{0},n_{d,\sigma'}^{0}\}$.

We define now the non-interacting kinetic energy functional $T[n_{c,i,\sigma},n_{d,\sigma'}]=Q[n_{c,i,\sigma},n_{d,\sigma'},U=0]$,
and the Exchange-correlation functional $E^{XC}=Q-T$, from which the exact Exchange-correlation
potential $V^{XC}$ is obtained by taking partial derivatives
\begin{eqnarray}
V_{c,i,\sigma}^{XC}[n_{c,i',\sigma'},n_{d,\sigma''}]\, & = & \,\frac{\partial E^{XC}}{\partial n_{c,i,\sigma}}\,\,\,\,\,\,\,\,\, i=1,{\cal M}\nonumber \\
V_{d,\sigma}^{XC}[n_{c,i',\sigma'},n_{d,\sigma''}]\, & = & \,\frac{\partial E^{XC}}{\partial n_{d,\sigma}}
\end{eqnarray}
We do not use a Hartree term in the definition of $E^{XC}$ because we have found no traces
of such a term in the analytic equations for the exact functionals. Therefore we have found it
useless for the purposes of the present discussion. We define the exact KS Hamiltonian as follows:
\begin{eqnarray}
H^{KS}= &  & \sum_{i,\sigma}(\epsilon_{c}\,+V_{c,i,\sigma}^{XC})\,\hat{n}_{c,i,\sigma}+\sum_{\sigma}(\epsilon_{d}+V_{d,\sigma}^{XC})\,\hat{n}_{d,\sigma}\nonumber \\
 &  & -t\,\sum_{i,\sigma}\,(\hat{c}_{i,\sigma}^{\dagger}\,\hat{d}_{\sigma}\,+\,\hat{d}_{\sigma}^{\dagger}\,\hat{c}_{i,\sigma}\,)-E_{dc}
\end{eqnarray}
where $E_{dc}$ is a double-counting term. Notice that the above procedure
allows to define functionals and KS Hamiltonians
for systems with a fractional electron number. However, the many-body Hamiltonian
in Eq. (3) commutes with the electron number operator $N_{\sigma}$.
Therefore the many-body Hamiltonian eigenstates must describe an integer
number of electrons, unless some degeneracy occurs.
We will see later on that the functional $Q$ has a polygonal shape,
so that the Exchange-correlation potentials jump by constants
at integer $N_{\sigma}$ values, which lead to ambiguous definitions
of the KS eigenvalues at integer $N_{\sigma}$. However, the total energies of
the ground- and excited-states of the KS Hamiltonian $E^{\alpha}$
($\alpha=0,1,...$) are continuous because the the jumps in the summations
over KS eigenvalues are counterbalanced by similar jumps in the double-counting
terms. The ground state energy $E^{0}$ of the exact KS Hamiltonian
and many-body Hamiltonians agree with each other by construction,
but this is not so for the total energies of the excited states of
both Hamiltonians, which are needed to construct the Green's functions.

The QP spectrum of the many-body Hamiltonian can be compared with
the KS and mean-field spectra by looking at the poles and residues
of the Green's functions $G(\omega)$, $G^{KS}(\omega)$ and $G^{MF}(\omega)$.
We define on this matter the many-body HOMO level as the QP peak which
is partially filled. The exact $G$ and $G^{KS}$ need not agree,
except for the pole describing the HOMO level. We will use the Lehmann
representation\cite{MahanBook} to compute $G$ for integer $N$-values.
In addition, the equations-of-motion method\cite{Lunquist}
yields a closed set of equations for $G$ for the single-site model.
This method nicely enables to extrapolate the $G$-poles to non-integer
electron numbers, and agrees with the results obtained using the Lehmann
representation for integer $N$. The mean field spectrum can be obtained
from the eigenvalues of the one-body mean-field Hamiltonian, or using
the equations-of-motion method for $G^{MF}$. The KS spectrum could also be obtained
from the eigenvalues of the one-body KS Hamiltonian. However, these
KS eigenvalues are discontinuous at $N$ integer so ambiguities in the ascription
of eigenvalues to QPs arise for integer $N$. It is therefore essential
to use the Lehmann representation as a guide.

We close this section by describing an alternative procedure which also allows
us to find exact results. If the exact ground state energy $E^0$ and occupations
$n^0_{i,\sigma}$ are found then the Schr\"odinger equation for the Kohn-Sham
hamiltonian can be inverted to find the exact exchange and correlation potential
corresponding to the ground state occupations
$v^{XC}_{i,\sigma}=V^{XC}_{i,\sigma}[n_{j,\sigma'}^0]$.
We note however that $v^{XC}_{i,\sigma}$ is not a functional, but rather corresponds to the
Exchange-correlation potential functional evaluated at the ground state occupations.
This procedure is simpler than the methodology described in
this section, but does not allow us to find functionals. Similar methods have
been employed by Baerend and coworkers\cite{Gritsenko96}, as well as by Helbig
and coworkers\cite{Helbig09} to find exact analytical or numerical expressions
for the Exchange-correlation potential of diatomic molecules in the dissociation limit.

\section{Single-site Anderson-Hubbard model}

The above methodology can be easily applied to the single-site Anderson-Hubbard
model (${\cal M}=0$), where only two occupations $\{n_{d,\uparrow},n_{d,\downarrow}\}$
are defined. The states in the Fock space of the single-site model
can be expressed using the number basis $|n_{d,\uparrow},n_{d,\downarrow}>$
as
\begin{equation}
|\,\phi>\,=\, a_{0}\,|\,0,0>\,+\, a_{\uparrow}\,|\,1,0>\,+\, a_{\downarrow}\,|\,0,1>\,+\, a_{2}\,|\,1,1>
\end{equation}
 The expectation value of the occupation numbers and the energy can
then be expressed as
\begin{eqnarray}
n_{d,\sigma}(\phi) & = & \frac{|a_{\sigma}|^{2}+|a_{2}|^{2}}{D}\nonumber \\
E(\phi) & = & \sum_{\sigma}\,\epsilon_{d}\, n_{d,\sigma}+\, U\,\frac{|a_{2}|^{2}}{D}
\end{eqnarray}
 where $D={|a_{0}|^{2}+|a_\uparrow|^{2}+|a_\downarrow|^{2}+|a_{2}|^{2}}$. The
simplest way to find $Q[n_{d,\uparrow},n_{d,\downarrow}]$ is as follows.
Solve first for some of the coefficients $a_{i}$ using the occupations
for the occupation numbers $n_{d,\sigma}$. Those coefficients are
then eliminated by inserting them back into the equation for $E(\phi)$.
The resulting expression is then minimized in terms of the remaining
coefficients. One must be careful though to choose coefficients which
are strictly non-zero in a given domain of $N$. In the present case,
it is best to solve for $a_{\sigma}$ since these are finite for all
$N=n_{d,\uparrow}+n_{d,\downarrow}$ different from $0,2$:
\begin{equation}
(1-N)\,(|a_{\uparrow}|^{2}+|a_{\downarrow}|^{2})=N\,|a_{0}|^{2}+(2-N)\,|a_{2}|^{2}
\end{equation}
 The resulting equation for
\begin{equation}
E(\phi)=\sum_{\sigma}\,\epsilon_{d}\, n_{d,\sigma}+\, U\,(N-1)\,\frac{|a_{2}|^{2}}{|a_{2}|^{2}-|a_{0}|^{2}}
\end{equation}
 is minimized as
\begin{equation}
\begin{array}{ccccl}
0<N<1 & \rightarrow & |a_{2}|^{2}=0\rightarrow & Q= & \epsilon_{d}\, N\\
1<N<2 & \rightarrow & |a_{0}|^{2}=0\rightarrow & Q= & \epsilon_{d}\, N+U\,(N-1)
\end{array}
\end{equation}

The above expressions for the exact functional $Q$ can be summarized
as
\begin{equation}
Q[n_{d,\uparrow},n_{d,\downarrow}]=\epsilon_{d}\, N\,+\, U\,(N-1)\,\theta(N-1)
\end{equation}
 where $\theta$ is the Heaviside step function. This expression gives
the correct ground state energy for a system with a target number
$N_{\uparrow}^{0},N_{\downarrow}^{0}$ of electrons
\begin{equation}
E^{0}=\epsilon_{d}\, N^{0}+U\,(N^{0}-1)\,\theta(N^{0}-1).
\end{equation}
 Note that this ground state energy is spin-degenerate.
Subtracting from $Q$ the non-interacting kinetic energy functional
$T[n_{d,\uparrow},n_{d,\downarrow}]=\epsilon_{d}\, N$, and taking
a functional derivative, we find the exact Exchange-correlation
potential $V_{d,\sigma}^{XC}=U\,\theta(N-1)$. The resulting KS Hamiltonian
\begin{equation}
H^{KS}=\sum_{\sigma}(\epsilon_{d}+U\,\theta(N-1))\,\hat{n}_{d,\sigma}-U\,\theta(N-1)
\end{equation}
provides the correct $E^{0}$ thanks to the double counting term
$U\,\theta(N^{0}-1)$. Notice that the KS eigenvalue jumps by $U$ exactly at
$N=1$, and is therefore ill-defined at that integer $N$-value\cite{Perdew81,Perdew82,Perdew83,Janak78}.
The density functional $Q$ has the correct trapezoidal shape\cite{Perdew82,Cohen08b}
as a function of $N_{\sigma}$, from which the right expression for
the chemical potential of the system can be obtained. Furthermore,
$Q$ shows flat-plane behavior when plotted as a function of the occupation
numbers, as displayed in Fig. 1. We note that Yang and coworkers established
some exact conditions on the shape of the exact energy functional,
from which they deduced such a flat-plane behavior\cite{Mori09}.
These conditions enabled them to draw an educated plot of the energy
functional of the hydrogen atom, which is very similar to our Fig.
1.

\begin{figure}
\includegraphics[width=0.9\columnwidth,height=5cm]{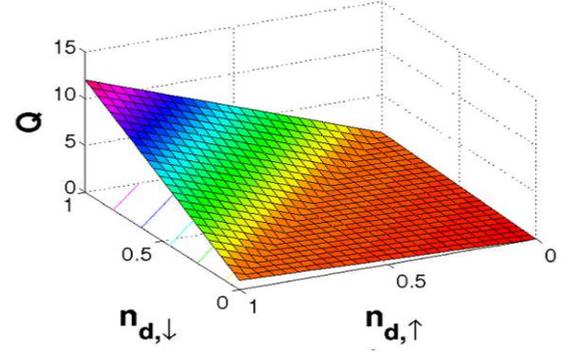} \caption{(Color online) Three-dimensional plot of the exact energy functional
$Q$ of the single-site Anderson-Hubbard model as a function of $(n_{d,\uparrow},n_{d,\downarrow})$,
for a value of $\epsilon_{d}=1$ and of $U=10$ (in arbitrary units). }
\end{figure}

We write down now the mean-field Hamiltonian of this model
\begin{equation}
H^{MF}=\sum_{\sigma}(\epsilon_{d}+U\, n_{d,-\sigma})\,\hat{n}_{d,\sigma}-U\, n_{d,\uparrow}\, n_{d,\downarrow}
\end{equation}
 where we have subtracted the conventional mean-field double-counting
term. The mean-field Hamiltonian gives the following estimate for
the energy of the system
\begin{eqnarray}
E^{MF} & = & \epsilon_{d}\, N+U\, n_{d,\uparrow}\, n_{d,\downarrow}=\epsilon_{d}\, N+U\,\frac{N^{2}-M^{2}}{4}\nonumber \\
M & = & n_{d,\uparrow}-n_{d,\downarrow}
\end{eqnarray}
 where the spin-degeneracy of the exact solution is lost. Note that
in the Hubbard and Anderson models every electron interacts only with
electrons of opposite spin. As a consequence, the mean-field theory
does not suffer from direct or exchange self-interaction effects.
However, because of the mean-field replacement of electron probabilities
by charge clouds, an electron of spin $\sigma$ interacts with a fraction
$n_{d,-\sigma}$ of electrons of opposite spin, instead of with a
full electron of opposite spin with probability $n_{d,-\sigma}$.
As a consequence, in the paramagnetic solution $M=0$, every electron
interacts artificially with a fraction $N/2$ of electrons of opposite
spin.
However, the mean-field ground state energy is minimized by the fully
spin polarized solutions $M=N$, in which case the spurious interaction
between opposite-spin charge clouds is avoided by a wrong mechanism
and, as a consequence, $E^{MF}=E^{0}$. In contrast, the interacting
piece of the exact KS Hamiltonian $U\,\theta(N-1)$ is
only activated if a full electron exists already in the system, and
therefore retains the full quantum behavior.

\begin{figure}
\includegraphics[width=0.8\columnwidth,height=5.5cm]{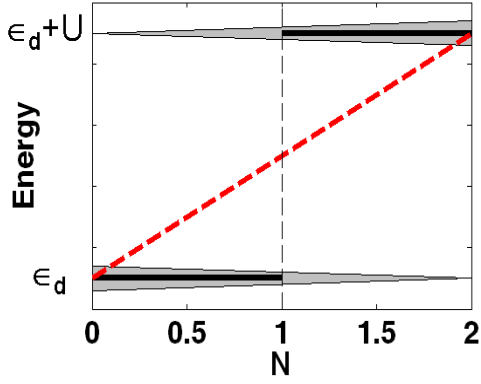} \caption{(Color online) Quasiparticle spectrum of the single-site Anderson-Hubbard
model for $M=0$. The shaded gray area shows the position of the many-body
poles, where the area is proportional to the weight of the peak. The
black solid line represents the location of the exact Kohn-Sham eigenstate.
The poles of the paramagnetic mean-field solution are shown with a
dashed red line. Energy units are arbitrary. }
\end{figure}

The exact many-body and mean-field QP spectrum are obtained
from the poles and weights of the retarded Green's function
\begin{eqnarray}
G_{d,\sigma}(\omega) & = & \frac{1-n_{d,-\sigma}}{\omega-\epsilon_{d}+i\delta}+\frac{n_{d,-\sigma}}{\omega-(\epsilon_{d}+U)+i\delta}\nonumber \\
G_{d,\sigma}^{MF}(\omega) & = & \frac{1}{\omega-(\epsilon_{d}+U\, n_{d,-\sigma})+i\delta}
\end{eqnarray}
The above equations can easily be obtained using the equations-of-motion
method and allow to extrapolate the QP spectrum to non-integer
values of $n_{d,\sigma}$ which, coupled to the spin degeneracy of
the total energy enable the exploration of different spin states.

We analyze first the paramagnetic state where $M=0$ and $n_{d,\sigma}=N/2$.
The exact KS Green's function is found by combining the equations-of-motion
method with the Lehmann representation for $N=0,1,2$. The following
formula summarizes the results
\begin{eqnarray}
G_{d,\sigma}^{KS}(\omega) & = & \frac{\theta(1-N)}{\omega-\epsilon_{d}+i\delta}+\frac{\theta(N-1)}{\omega-(\epsilon_{d}+U)+i\delta}
\end{eqnarray}
and extrapolates them to non-integer-$N$-values. The many-body,
exact DFT and mean field Green's function are shown in Fig. 2 as a
function of $N$. The many-body Green's function has two poles, which
can be viewed as the ancestors of the lower and upper Hubbard bands
of the Hubbard and Anderson models\cite{HewsonBook}. The position
of these two poles depends neither on the occupation nor on the spin
of the system. They are separated exactly by an energy $U$ and their
weight shifts smoothly from one peak to the other as $N$ increases.
Because exact DFT is a single-particle theory, its Green's function
yields a single peak per KS eigenvalues, whose weight equals one.
A remarkable exception happens at $N=1$, where the KS eigenvalue show an abrupt
change from $\epsilon_{d}$ to $\epsilon_{d}+U$. Notice that both
eigenvalues contribute at $N=1$ with equal weight. The $N\rightarrow1^{-}$
($N\rightarrow1^{+}$) KS eigenvalue exactly agrees with the many-body
lower (upper) Hubbard band precursor. This non-trivial result allows
to draw an important conclusion: even if KS eigenvalues show a jump
at integer electron number values, the eigenvalues at both sides of
the given $N$ contribute to the QP spectrum. To summarize, the positions
and weights of the exact KS and many-body peaks coincide for integer
numbers $N$, showing how exact DFT keeps the quantum nature of the
electrons in spite of being a one-body theory. The abrupt shift at
$N=1$ can be viewed as the way that exact DFT uses to retain that
quantum nature: if there is less than one electron at the site, then
there is no Coulomb interaction because the electrons does not interact
with itself. If there is more than one electron, then the Coulomb
interaction between point particles of opposite spin is activated,
rising the energy by $U$. Notice that many-body and exact DFT agree
on the value of HOMO level, which is also equal to the chemical potential
$\mu$, defined as the derivative of the total energy with respect
to the particle number\cite{Perdew83,Sham83}.

\begin{figure}
\includegraphics[width=0.9\columnwidth,height=9cm]{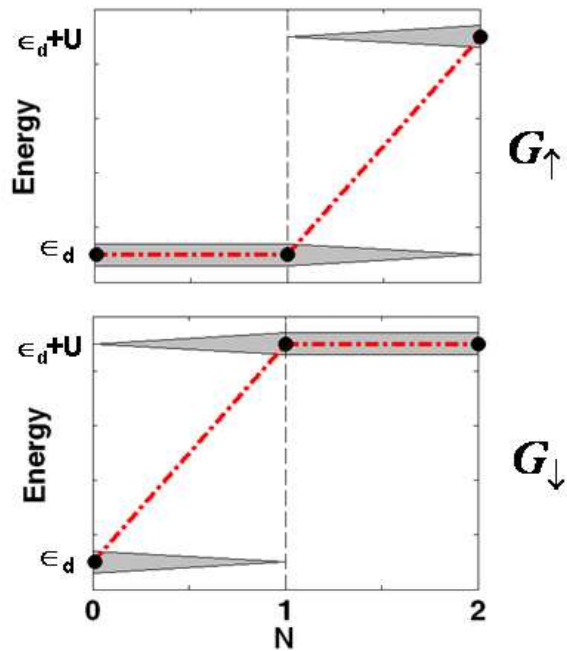} \caption{(Color online) Quasi-particle spectrum of the single-site Anderson-Hubbard
model for a maximally spin-up polarized case. (a) and (b) show the
poles of the spin-up and spin-down Green's functions, respectively.
The shaded gray area represents the many-body quasi-particle, where
the width is proportional to the peak weight. The black thick dot
represents the location of the exact Kohn-Sham eigenstates. The poles
of the spin-polarized mean-field solution are shown with a dashed
red line. Energy units are arbitrary. }
\end{figure}

We must remember however that in this quantum system only states with
integer electron numbers $N_{\sigma}=0,1$ are meaningful. Therefore
for $N=1$, the system must contain a full electron with either spin
up or down. We therefore turn now to analyze a maximally spin-up polarized.
The up- and down-spin Green's functions are different now, as is apparent
from the Lehmann representation at $N=1$, where we take $|1,0>$
as the ground state,
\begin{eqnarray}
G_{d,\uparrow} & = & \frac{|<0,0|\,\hat{c}_{d,\uparrow}\,|1,0>|^{2}}{\omega+E_{1}-E_{0}+i\,\delta}\nonumber \\
G_{d,\downarrow} & = & \frac{|<1,1|\,\hat{c}_{d,\downarrow}^{\dagger}\,|1,0>|^{2}}{\omega+E_{1}-E_{2}+i\,\delta}
\end{eqnarray}
 where $E_{N}$ denote the total ground state energy for $N=0,1,2$.
We determine now the many-body and mean field QP spectra for fractional
occupation numbers using Eqs. (16). To determine $G^{KS}$ correctly
for integer $N$ we use the Lehmann representation. The following
formula extrapolates $G^{KS}$ to non-integer $N$-values
\begin{eqnarray}
G_{d,\uparrow}^{KS}(\omega) & = & \frac{\theta(1-N+\delta)}{\omega-\epsilon_{d}+i\delta}\nonumber \\
G_{d,\downarrow}^{KS}(\omega) & = & \frac{\theta(N-1+\delta)}{\omega-(\epsilon_{d}+U)+i\delta}
\end{eqnarray}
The different spectra are shown in Fig. 3. As before, the exact KS eigenvalues
agree with the many-body QP for integer $N$. The mean-field states
have a closer resemblance to the many-body QP, although clear differences
still exist, whose origin is traced back to the static correlation
error. As a closing remark, we note that the GW approximation cures
these mean-field artifacts for the present case as shown by Romaniello
and coworkers\cite{Romaniello09}.

\section{double-site Anderson model}

The model in the previous section has allowed to show how the exact
density functional retains the quantum nature of electrons in a single
atom and therefore avoids the static correlation error brought about
by mean field theory. We wish to address in this section how the exact
functional avoids also the delocalization error in a strongly-correlated
model containing two sites. We show that the exact KS Hamiltonian
provides a correct description of the atomic limit of the correlated
model. The discussion is centered in the Anderson model, but our conclusions
can be also applied to the double-site Hubbard model, which is solved
in Appendix A. Notice that the many-body QP spectrum of
the full Anderson model is much more complex than that of the single-site
model discussed above and, in addition to the lower and upper Hubbard
bands it develops a Kondo resonance in the Kondo regime. We therefore
wish to explore now whether exact DFT could describe this more convoluted
QP spectra. Finally, notice that the obtained energy functional $Q$
is spin-degenerate, in contrast to the full Anderson and Hubbard models,
where this degeneracy is absent. It is therefore interesting to check
whether exact DFT lifts the spin-degeneracy for more realistic models.

\begin{figure}
\includegraphics[width=0.75\columnwidth,height=5.5cm]{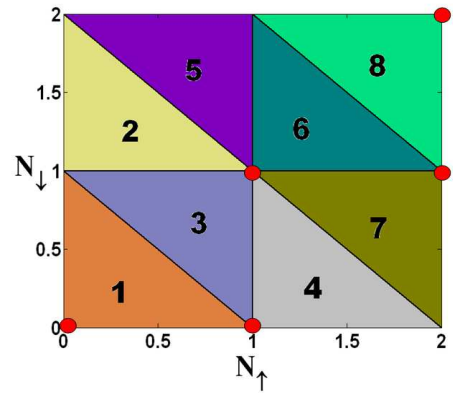} \caption{(Color online) The eight pieces in the $(N_{\uparrow},N_{\downarrow})$-plane
which must be used to perform the constrained minimization procedure
leading to the exact $Q$-functional of the double-site Anderson and
Hubbard models. The thick dots indicate the positions where $G$ and
$G^{KS}$ are evaluated. }
\end{figure}

We describe here the exact DFT solution of the double-site Anderson,
which corresponds to taking ${\cal M}=1$ in Eq. (1), and can also
be solved analytically. The number basis $\{|n_{c,\sigma}>,|n_{d,\sigma}>\}$
of the Fock space is spanned by sixteen states, which renders the
minimization task of finding $Q[n_{c,\sigma},n_{d,\sigma}]$ asymptotically
harder. We have found that the electron number plane $(N_{\uparrow},N_{\downarrow})$
is split into eight pieces as shown in Fig. 4, such that in each piece
only a subset of the wave-function coefficients is different from
zero. As a consequence, the minimization task has to be performed
separately for each of those pieces. The $Q$-functional has again
a polygonal shape. After lengthy algebra, the following expressions
for the $Q$-functional in the symmetric case $\epsilon_{d}+U/2=\epsilon_{c}$
can be written: \begin{widetext}
\begin{equation}
\begin{array}{lll}
F_{1} & = & -\left[\,\sqrt{n_{c,\uparrow}\, n_{d,\uparrow}}+\sqrt{n_{c,\downarrow}\, n_{d,\downarrow}}\,\right]\\
\\
F_{2} & = & -\left[\,\sqrt{(1-n_{c})\,(1-n_{d})}+2\, x\,\left(\sqrt{n_{c,\uparrow}-x^{2}}+\sqrt{n_{d,\uparrow}-x^{2}}\,\right)\,\right]+U\, x^{2},\,\,\, x=\frac{1}{2}\,\sqrt{N_{\uparrow}\,\left(1-\frac{U}{\sqrt{U^{2}+16\, t^{2}}}\right)}\\
\\
F_{3} & = & -\frac{1}{2}\,\left[\,\sqrt{(1+m_{c}-n_{d})\,(1+m_{d}-n_{c})}+\sqrt{(1-m_{c}-n_{d})\,(1-m_{d}-n_{c})}\right]+\frac{N-1}{4}\,\left(U-\sqrt{U^{2}+64\, t^{2}}\,\right)\\
\\
F_{4} & = & -\left[\,\sqrt{(1-n_{c})\,(1-n_{d})}+2\, x\,\left(\sqrt{n_{c,\downarrow}-x^{2}}+\sqrt{n_{d,\downarrow}-x^{2}}\,\right)\,\right]+U\, x^{2},\,\,\, x=\frac{1}{2}\,\sqrt{N_{\downarrow}\,\left(1-\frac{U}{\sqrt{U^{2}+16\, t^{2}}}\right)}\\
\\
F_{5} & = & -\left[\,\sqrt{(n_{c}-1)\,(n_{d}-1)}+2\, x\,\left(\sqrt{1-n_{c,\downarrow}-x^{2}}+\sqrt{1-n_{d,\downarrow}-x^{2}}\,\right)\right]+U\,(n_{d}-1+x^{2}),\\
 &  & x=\frac{1}{2}\,\sqrt{(2-N_{\downarrow})\,\left(1-\frac{U}{\sqrt{U^{2}+16\, t^{2}}}\right)}\\
\\
F_{6} & = & -\frac{1}{2}\left[\,\sqrt{(n_{d}-m_{c}-1)\,(n_{c}-m_{d}-1)}+\sqrt{(n_{d}+m_{c}-1)\,(n_{c}+m_{d}-1)}\right]+\frac{3-N}{4}\,\left(U-\sqrt{U^{2}+64\, t^{2}}\,\right)+U\,(n_{d}-1)\\
\\
F_{7} & = & -\left[\,\sqrt{(n_{c}-1)\,(n_{d}-1)}+2\, x\,\left(\sqrt{1-n_{c,\uparrow}-x^{2}}+\sqrt{1-n_{d,\uparrow}-x^{2}}\,\right)\right]\,+U\,(n_{d}-1+x^{2})\\
 &  & x=\frac{1}{2}\,\sqrt{(2-N_{\uparrow})\,\left(1-\frac{U}{\sqrt{U^{2}+16\, t^{2}}}\right)}\\
\\
F_{8} & = & -\left[\,\sqrt{(1-n_{c,\uparrow})\,(1-n_{d,\uparrow})}+\sqrt{(1-n_{c,\downarrow})\,(1-n_{d,\downarrow})}\,\right]+U\,(n_{d}-1)
\end{array}
\end{equation}
 \end{widetext} where we have defined a different
\[
F_{a}=\frac{Q-n_{c}\,\epsilon_{c}-n_{d}\epsilon_{d}}{2|t|}
\]
 for each of the eight $a$-zones depicted in Fig. 4. We also use
the site-occupations and moments as $n_{i},\, m_{i}=n_{i,\uparrow}\pm n_{i,\downarrow}$
with $i=c,d$. The full expressions for $F_{a}$ are shown in Appendix
B. Simplified expression, valid along the line $N_{\uparrow}+N_{\downarrow}=2$
are also provided in the appendix. Finally, the ground-state energy
$E^{0}$ for given electron numbers $N_{\sigma}^{0}$ is found by
minimizing $Q$ with the constraints $N_{\sigma}^{0}=n_{c,\sigma}+n_{d,\sigma}$.
To simplify the notation, energies will be measured in units of $|t|$,
and the energy origin will be chosen at $\epsilon_{c}$ from now on.

\begin{figure}
\includegraphics[width=0.9\columnwidth]{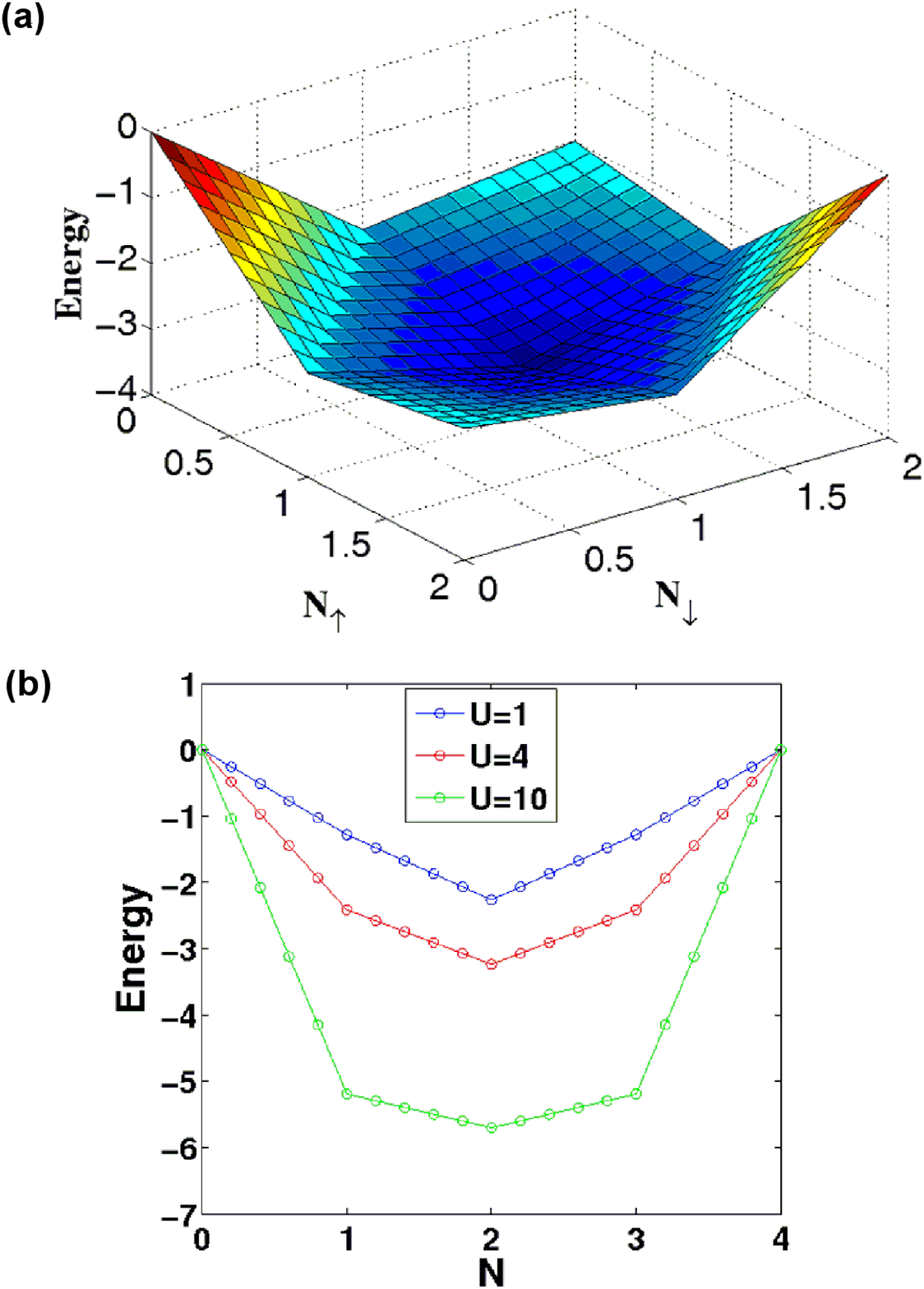} \caption{(Color online) (a) three-dimensional plot of the ground state energy
as a function of $(N_{\uparrow},N_{\downarrow})$ for the symmetric
case with $U=4$. (b) Ground state energy along the paramagnetic line
in the symmetric case for several $U$-values. Energies are given
in units of $|t|$. }
\end{figure}

We find that $Q$ is spin-degenerate only in regions 1 and 8 of Fig.
4, where $N$ is smaller than $1$, or bigger than $3$. However,
we find that the spin-degeneracy is lifted if $1<N<3$, because here
the interplay between kinetic energy and Coulomb interactions is more
convoluted. The minima of $Q$ and $E^{0}$ occur now along the paramagnetic
line $M=N_{\uparrow}-N_{\downarrow}=0$ regardless of the value of
the on-site energy $\epsilon_{d}$, and of $U$. This is shown in
Fig. 5(a), where the ground-state energy is plotted in the $(N_{\uparrow},N_{\downarrow})$-plane
for the symmetric case and $U=4$. Here the characteristic polygonal
shape as well as the presence/absence of spin degeneracies in the
different regions are apparent. The position of the absolute minimum
of $E^{0}$ along the paramagnetic line in contrast does depend on
$\epsilon_{d}$ and on $U$. For the symmetric case, $\epsilon_{d}+U/2=0$,
the minimum is placed at $N=2$. Fig. 5(b) shows $E^{0}$ as a function
of $N$ along the paramagnetic line for the symmetric case and for
several values of $U$ which cover the weak-, intermediate- and strong-coupling
regimes of the model. The chemical potential and the energy value
of the HOMO are given by the slope of these curves. They exhibit the
expected discontinuous behavior at integer values of $N$\cite{Perdew82,Perdew83,Mori09}.
Finally, it can be checked that this $Q$ functional renders the correct
atomic limit by taking explicitly $t\rightarrow0$ in Eq. (17). As
a consequence the exact $Q$ functional is free from the delocalization
error of mean-field theory.
The analytic expressions for $Q$ enable to find the exact Exchange-correlation
potentials for the two-site model $(V_{c,\sigma}^{XC}$, $V_{d,\sigma}^{XC})$.
Notice that these potentials keep the full non-local dependence on occupations, because the potential
at a given site $(i,\sigma)$ depends on all the densities $n_{j,\sigma'}$.
In contrast, it is very difficult to determine accurately the non-local
terms by a numerical solution of this model, or by extending the Bethe
ansatz LDA approach. We define the exact KS Hamiltonian
for this double-site Anderson model as
\begin{equation}
H_{KS}=\sum_{i=(c,d),\sigma}(\epsilon_{i}\,+V_{i,\sigma}^{XC})\,\hat{n}_{i,\sigma}-t\,\sum_{\sigma}\,(\hat{c}_{\sigma}^{\dagger}\,\hat{d}_{\sigma}\,+\,\hat{d}_{\sigma}^{\dagger}\,\hat{c}_{\sigma}\,)-H_{dc}
\end{equation}

\begin{figure}[t]
 \includegraphics[width=1\columnwidth]{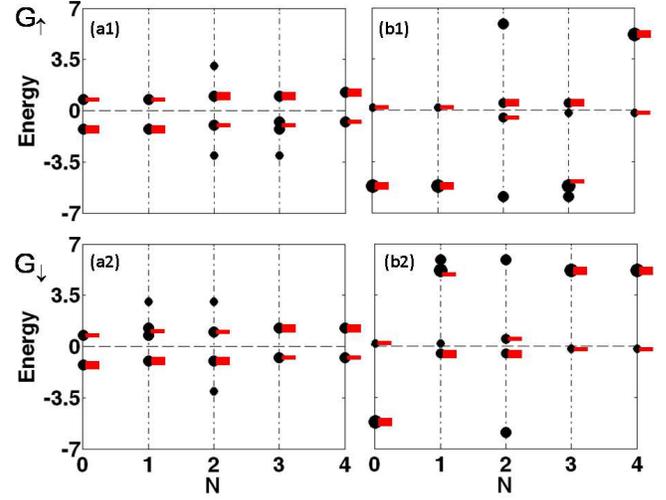} \caption{(Color online) Quasi-particle spectrum of the symmetric double-site
Anderson model as a function of the electron number $N$, computed
at the thick dots shown in Fig. 4. Green's functions poles for $U=1$
((a1) and (a2)) and $U=10$ ((b1) and (b2)). The upper and lower panels
show $G_{d,\uparrow}$ and $G_{d,\downarrow}$, respectively. The
energies of the many-body quasi-particles are shown as black dots
whose width is proportional to the quasi-particle weight. the Kohn-Sham
eigenstates are displayed as red dashes. The position of the one-particle
HOMO level is marked by a thicker dash. Energies are given in units
of $|t|$. }
\end{figure}

This Hamiltonian only has two KS eigenvalues per spin for all values
of the physical parameters, which are discontinuous at integer $N_{\sigma}$-values.
In other words, the numerical values of the KS eigenvalues are constant within
each of the eight regions in Fig. 4, but differ from region to region.
We compare now the KS eigenvalues with the exact many-body QP spectrum
extracted from the poles of the many-body Green's function at the
impurity's position. Notice again that only integer electron numbers
$N=0,1,2,3,4$ have a physical meaning. For $N=1$, the system contains
a single electron which must have either spin up or down. If a ground
state wave-function with spin up $|\Psi_{1,\uparrow}>$ is chosen,
then the spin-up and -down Green's functions are different,
\begin{eqnarray}
G_{d,\uparrow} & = & \frac{|<\Psi_{2,\uparrow,\uparrow}|\, c_{d,\uparrow}^{\dagger}\,|\Psi_{1,\uparrow}>|^{2}}{\omega+E_{1,\uparrow}-E_{2,\uparrow,\uparrow}+i\,\delta}+\frac{|<\Psi_{0}|\, c_{d,\uparrow}\,|\Psi_{1,\uparrow}^{1}>|^{2}}{\omega+E_{1,\uparrow}^{1}-E_{0}+i\,\delta}\nonumber \\
G_{d,\downarrow} & = & \sum_{n}\frac{|<\Psi_{2}^{n}|\, c_{d,\downarrow}^{\dagger}\,|\Psi_{1,\uparrow}>|^{2}}{\omega+E_{1,\uparrow}^{1}-E_{2}^{n}+i\,\delta}
\end{eqnarray}
 where the summation runs over all spin-0 states with $N=2$, and
$\Psi_{2,\uparrow,\uparrow}$ indicate the spin-1 $N=2$ state. Similar
words can be said for $N=3$. Romaniello and coworkers\cite{Romaniello09}
have compared the spectrum of many-body QPs of this model with the
poles of Green functions evaluated either in the GW approximation,
or including vertex corrections. They have shown that the mean-field
static correlation error is amended. However, even inclusion of vertex
corrections does not allow to recover the QP spectrum in the atomic
limit, showing how hard is to fully get rid of the delocalization
error.

The exact KS Green's function can be computed using the equations-of-motion method
giving rise to the following expression
\begin{equation}
G_{d,\sigma}^{KS}=\frac{\omega-\epsilon_{c}-V_{c,\sigma}^{XC}}{(\omega-\epsilon_{c}-V_{c,\sigma}^{XC})\,(\omega-\epsilon_{d}-V_{d,\sigma}^{XC})-|t|^{2}}
\end{equation}

This formula must be guided by the results obtained from the Lehmann
representation at integer $N$.
We compare now the poles of the many-body and exact KS Green's functions
by evaluating $V^{XC}$ at the points in the $(N_{\uparrow},N_{\downarrow})$
path shown in Fig. 4. This correspond to a paramagnetic solution for
$N=2$ and a spin-up state for $N=1,3$. Fig. 6 shows the poles of
$G$ and $G^{KS}$ as a function of the electron number $N$ for a
symmetric case, and for values of $U$ in the weak- and strong-coupling
regimes. The figure also shows which of the KS eigenvalues
corresponds to the HOMO level. Notice that the exact many-body and KS spectra
closely match for values of $U$ not only in the weakly-correlated,
but also in the strongly-correlated regimes. However, extra many-body
peaks appear at $N=1,2,3$, which are not provided by the exact KS
Hamiltonian. In contrast, the number of many-body and KS QPs
is the same for $N=0,4$ because an electron added to an empty system
or a hole added to a fully occupied system can not Coulomb-interact
with anything. Occupation $N=2$ corresponds to the strongly-correlated
Kondo regime if $U$ is large, which is the case shown in Fig. 6(b).
Here the many-body QP spectrum has four poles. These can be classified
into two sets of peaks placed symmetrically about the zero-energy
line. The first set is located around $\pm U/2$. The two peaks are
separated by an energy of order $U$ and correspond to the upper and
lower Hubbard bands. The second set develops into the Kondo resonance
for more realistic models where ${\cal M}$ is made large. The KS
spectrum has only two QP, which agree with the two Kondo-like many-body
QPs. In other words, the KS spectrum shows no trace now of the lower
and upper Hubbard band precursors. For $N=1$, the up-spin KS spectrum
matches the many-body spectrum because the many-body phase-space for
adding a spin-up electron or hole is very limited. However, the many-body
phase-space for the addition of a spin-down electron is larger, which
renders additional many-body spin-down quasi-electron peaks. Similar
words can be said for $N=3$, where additional spin-up quasi-hole
peaks are apparent. Notice in any case that the many-body and KS HOMO
levels always agree with each other.

\section{Conclusions}

We have presented analytic expressions for the exact density functionals
of several simple models of strongly correlated electrons, from which
we have obtained the exact ground-state energy. Those analytic expressions
have allowed us to write down the full non-local dependence of $V^{XC}$ and KS
Hamiltonians on the occupations. We have computed the exact KS eigenvalues and compared them
with the true many-body QP, as obtained from the poles
of the Green's functions. We have shown with explicit examples that
exact DFT preserves the quantum nature of electron-electron interactions,
as opposed to mean-field theory and improvements over it as the GW
approximation. It is also superior to more sophisticated perturbative
approximations including vertex corrections. The exact functionals
do not show any trace of self-interaction, static correlation, or
delocalization errors.

We have found that the KS eigenvalues spectrum agrees to a large
extent, but not fully, with the exact many-body spectrum. This is
to say that all KS eigenvalues agree with some of the many-body QPs.
However, the many-body QP spectrum is richer because the phase-space
for addition of quasi-electrons or quasi-holes is larger. The exact
functional only warrants the correct position of the HOMO level, while
in general other KS eigenvalues may or may not agree with the exact
many-body QP. Remarkably, we have found that the KS spectrum most
possibly describes the Kondo peak in the Kondo regime. However, it
is quite plausible that it won't contain either the lower and upper
Hubbard bands, or both. Exact DFT has similarities with the Renormalized
Perturbation Theory proposed some time ago by Hewson\cite{Hewson93}.
The perturbative expansion shown in Eq. (2) would possibly describe
the full many-body spectrum with simple approximations for the self-energy.

\begin{acknowledgments}
J. F. would like to acknowledge conversations with V. M. Garc\'{\i}a-Su\'arez,
J. H. Jefferson, C. J. Lambert and M. A. R. Osorio, as well as help
with one equation from I. Zapata. K. Burke pointed out the relevance
of the results in Ref. \onlinecite{Romaniello09}. The research
presented here was funded by the Spanish MICINN through the grants
FIS2009-07081 and PR2009-0058, as well as by the Marie Curie network
nanoCTM.
\end{acknowledgments}

\appendix
\begin{widetext}

\section{Double-site Hubbard model}
We use the following notation for the hamiltonian of the double-site
Hubbard model
\begin{eqnarray*}
\hat{H}= &  & \sum_{\sigma}\,\epsilon_{0}\,(\hat{n}_{1,\sigma}\,+\hat{n}_{2,\sigma})-\, t_{0}\,\sum_{\sigma}\,(\,\hat{c}_{1,\sigma}^{\dagger}\,\hat{c}_{2,\sigma}\,+\,\hat{c}_{2,\sigma}^{\dagger}\,\hat{c}_{1,\sigma}\,)\,+\, U\,\sum_{i=1,2}\hat{n}_{i,\uparrow}\,\hat{n}_{i,\downarrow}
\end{eqnarray*}
The expressions for exact density functional are quite similar to
those of the double-site Anderson model,
\[
\begin{array}{lll}
F_{1} & = & -\left[\,\sqrt{n_{1,\uparrow}\, n_{2,\uparrow}}+\sqrt{n_{1,\downarrow}\, n_{2,\downarrow}}\,\right]\\
\\
F_{2} & = & -\left[\,\sqrt{(1-n_{1}+x^{2}-y^{2})\,(1-n_{2}-x^{2}+y^{2})}+(x+y)\,(\sqrt{n_{1,\uparrow}-x^{2}}+\sqrt{n_{2,\uparrow}-y^{2}})\,\right]+U\,(x^{2}+y^{2})\\
\\
F_{3} & = & -\left[\,\sqrt{(n_{1,\uparrow}-x^{2}-z^{2})\,(1-n_{1}-n_{2,\downarrow}+x^{2}+z^{2})}+\sqrt{(n_{2,\downarrow}-y^{2}-z^{2})\,(1-n_{1,\uparrow}-n_{2}+y^{2}+z^{2})}\right.\\
\,\,\,\,\,\, &  & +\left.(x+y)\,\left(z+\sqrt{n_{1}+n_{2}-1-x^{2}-y^{2}-z^{2}}\right)\right]+U\,(x^{2}+y^{2})\\
\\
F_{4} & = & -\left[\,\sqrt{(1-n_{1}+x^{2}-y^{2})\,(1-n_{2}-x^{2}+y^{2})}+(x+y)\,(\sqrt{n_{1,\downarrow}-x^{2}}+\sqrt{n_{2,\downarrow}-y^{2}})\,\right]+U\,(x^{2}+y^{2})\\
\\
F_{5} & = & -\left[\,\sqrt{(1-n_{1}+x^{2}-y^{2})\,(1-n_{2}-x^{2}+y^{2})}+(x+y)\,(\sqrt{1-n_{1,\downarrow}-y^{2}}+\sqrt{1-n_{2,\downarrow}-x^{2}})\,\right]\\
 &  & \,\,\,\,\,\,\,\,\,\,\,\,\,\,\,\,\,+U\,(n_{1}+n_{2}-2+x^{2}+y^{2})\\
\end{array}
\]
\[
\begin{array}{lll}
F_{6} & = & -\left[\,\sqrt{(n_{1,\downarrow}-1+y^{2}+z^{2})\,(2-n_{1}-n_{2,\uparrow}-y^{2}-z^{2})}+\sqrt{(n_{2,\uparrow}-1+x^{2}+z^{2})\,(2-n_{1,\downarrow}-n_{2}-x^{2}-z^{2})}\right.\\
 &  & \,\,\,\,\,\,\,\,\,\,\,\,\,\,+\left.(x+y)\,\left(z+\sqrt{3-n_{1}-n_{2}-x^{2}-y^{2}-z^{2}}\right)\right]+U\,(n_{1}+n_{2}-2+x^{2}+y^{2})\\
\\
F_{7} & = & -\left[\,\sqrt{(1-n_{1}+x^{2}-y^{2})\,(1-n_{2}-x^{2}+y^{2})}+(x+y)\,(\sqrt{1-n_{1,\uparrow}-y^{2}}+\sqrt{1-n_{2,\uparrow}-x^{2}})\,\right]\\
 &  & \,\,\,\,\,\,\,\,\,\,\,\,\,\,\,\,\,+U\,(n_{1}+n_{2}-2+x^{2}+y^{2})\\
\\
F_{8} & = & -\left[\,\sqrt{(1-n_{1,\uparrow})\,(1-n_{2,\uparrow})}+\sqrt{(1-n_{1,\downarrow})\,(1-n_{2,\downarrow})}\,\right]+U\,(n_{1}+n_{2}-2)
\end{array}
\]
where the functionals $F_{a}$ are defined as
\[F_{a}=\frac{<\hat{H}>-(n_{1}+n_{2})\,\epsilon_{0}}{2|t_{0}|}
\]
and where $Q$ is found by minimizing $F_a$ with respect to $x$ and $y$. Along the line $N_{\uparrow}+N_{\downarrow}=2$, the formulae for
the $F$-functional can be simplified as follows: 
\begin{eqnarray*}
F & = & -\left(\sqrt{n_{1,\downarrow}-x^{2}}+\sqrt{1-n_{2,\uparrow}-x^{2}}\right)\,\left(x+\sqrt{1-n_{1}+x^{2}}\right)+U\,(1-n_{1}+2x^{2})\\
\\
 & = & -\left(\sqrt{n_{1,\uparrow}-x^{2}}+\sqrt{1-n_{2,\downarrow}-x^{2}}\right)\,\left(x+\sqrt{1-n_{1}+x^{2}}\right)+U\,(1-n_{1}+2x^{2})
\end{eqnarray*}
 where the first equation is obeyed if $N_{\uparrow}>1$, $N_{\downarrow}<1$,
and vice versa. $Q$ is now obtained by minimizing the above equation
with respect to $x$.

\section{Double-site Anderson model}

The full expressions for $F_{a}$ are as follows:
\begin{equation}
\begin{array}{lll}
F_{1} & = & -\left[\,\sqrt{n_{c,\uparrow}\, n_{d,\uparrow}}+\sqrt{n_{c,\downarrow}\, n_{d,\downarrow}}\,\right]\\
\\
F_{2} & = & -\left[\,\sqrt{(1-n_{c}+x^{2}-y^{2})\,(1-n_{d}-x^{2}+y^{2})}+(x+y)\,(\sqrt{n_{c,\uparrow}-x^{2}}+\sqrt{n_{d,\uparrow}-y^{2}})\,\right]+U\, y^{2}\\
\\
F_{3} & = & -\left[\,\sqrt{(n_{c,\uparrow}-x^{2}-z^{2})\,(1-n_{c}-n_{d,\downarrow}+x^{2}+z^{2})}+\sqrt{(n_{d,\downarrow}-y^{2}-z^{2})\,(1-n_{c,\uparrow}-n_{d}+y^{2}+z^{2})}\right.\\
 &  & \,\,\,\,\,\,\,\,\,\,\,\,\,+\left.(x+y)\,\left(z+\sqrt{N-1-x^{2}-y^{2}-z^{2}}\right)\right]+U\, y^{2}\\
\\
F_{4} & = & -\left[\,\sqrt{(1-n_{c}+x^{2}-y^{2})\,(1-n_{d}-x^{2}+y^{2})}+(x+y)\,(\sqrt{n_{c,\downarrow}-x^{2}}+\sqrt{n_{d,\downarrow}-y^{2}})\,\right]+U\, y^{2}\\
\\
F_{5} & = & -\left[\,\sqrt{(1-n_{c}+x^{2}-y^{2})\,(1-n_{d}-x^{2}+y^{2})}+(x+y)\,(\sqrt{1-n_{c,\downarrow}-y^{2}}+\sqrt{1-n_{d,\downarrow}-x^{2}})\,\right.+U\,(n_{d}-1+x^{2})\\
\\
F_{6} & = & -\left[\,\sqrt{(n_{c,\downarrow}-1+y^{2}+z^{2})\,(2-n_{c}-n_{d,\uparrow}-y^{2}-z^{2})}+\sqrt{(n_{d,\uparrow}-1+x^{2}+z^{2})\,(2-n_{c,\downarrow}-n_{d}-x^{2}-z^{2})}\right.\\
 &  & \,\,\,\,\,\,\,\,\,\,\,\,\,\,\,\,\,+\left.(x+y)\,\left(z+\sqrt{3-N-x^{2}-y^{2}-z^{2}}\right)\right]+U\,(n_{d}-1+x^{2})\\
\\
F_{7} & = & -\left[\,\sqrt{(1-n_{c}+x^{2}-y^{2})\,(1-n_{d}-x^{2}+y^{2})}+(x+y)\,(\sqrt{1-n_{c,\uparrow}-y^{2}}+\sqrt{1-n_{d,\uparrow}-x^{2}})\,\right]+U\,(n_{d}-1+x^{2})\\
\\
F_{8} & = & -\left[\,\sqrt{(1-n_{c,\uparrow})\,(1-n_{d,\uparrow})}+\sqrt{(1-n_{c,\downarrow})\,(1-n_{d,\downarrow})}\,\right]+U\,(n_{d}-1)
\end{array}
\end{equation}
Q is again found by minimizing $F_a$ with respect to $x$ and $y$. Along the line $N_{\uparrow}+N_{\downarrow}=2$, the formulae for
the $F$-functional can be simplified, and read as follows: 
\begin{eqnarray*}
F & = & -\left(\sqrt{n_{c,\downarrow}-x^{2}}+\sqrt{1-n_{d,\uparrow}-x^{2}}\right)\,\left(x+\sqrt{1-n_{c}+x^{2}}\right)+U\,(n_{d}-1+x^{2})\\
 & = & -\left(\sqrt{n_{c,\uparrow}-x^{2}}+\sqrt{1-n_{d,\downarrow}-x^{2}}\right)\,\left(x+\sqrt{1-n_{c}+x^{2}}\right)+U\,(n_{d}-1+x^{2})
\end{eqnarray*}
where the first equation is obeyed if $N_{\uparrow}>1$, $N_{\downarrow}<1$,
and vice versa. $Q$ is now obtained by minimizing the above equation
with respect to $x$.

\end{widetext}

\section{{\cal M}-site spinless fermion model}

We show in this appendix the exact DFT solution of the double-site
spinless fermion model, which corresponds to taking ${\cal M}=1$
and discarding the spin index in Eq. (1),
\begin{equation}
\hat{H}=\epsilon_{c}\,\hat{n}_{c}+\epsilon_{d}\,\hat{n}_{d}-t(\hat{c}^{\dagger}\,\hat{d}+\hat{d}^{\dagger}\,\hat{c})+U\hat{n}_{d}\,\hat{n}_{c}
\end{equation}
 We use a variational wave function of the form
\begin{equation}
|\,\phi>\,=\, a_{0}\,|\,0,0>\,+\, a_{c}\,|\,1,0>\,+\, a_{d}\,|\,0,1>\,+\, a_{cd}\,|\,1,1>
\end{equation}
 to find explicit formulae for the expectation values of $\hat{H}$
and $\hat{n}_{c}$, $\hat{n}_{d}$ as a function of the parameters
$a_{i}$,
\begin{eqnarray}
<\hat{H}> & = & \epsilon_{c}\, n_{c}+\epsilon_{d}\, n_{d}-2\, t\cos\varphi\,\,\frac{|a_{c}|\,|a_{d}|}{D}\nonumber \\
<\hat{n}_{c}> & = & \frac{|a_{c}|^{2}\,+\,|a_{cd}|^{2}}{D}\nonumber \\
<\hat{n}_{d}> & = & \frac{|a_{d}|^{2}\,+\,|a_{cd}|^{2}}{D}\\
D & = & |a_{0}|^{2}+|a_{c}|^{2}\,+|a_{d}|^{2}+|a_{cd}|^{2}\nonumber
\end{eqnarray}
We solve for $|a_{c}|$, $|a_{d}|$ in the above equations for $n_{c,d}$
and substitute the result back in the equation for $<H>$. This yields
the following expressions for $<\hat{H}>-\epsilon_{c}\,<\hat{n}_{c}>-\epsilon_{d}\,<\hat{n}_{d}>$
\begin{eqnarray}
 & -2\, t\, cos\varphi\,\,\frac{\sqrt{n_{d}|a_{0}|^{2}+(n_{c}-1)|a_{cd}|^{2}}\,\,\sqrt{n_{c}|a_{0}|^{2}+(n_{d}-1)|a_{cd}|^{2}}}{|a_{0}|^{2}-|a_{cd}|^{2}}\nonumber \\
 & -2\, t\, cos\varphi\,\,\frac{\sqrt{(1-n_{c})|a_{cd}|^{2}-n_{d}|a_{0}|^{2}}\,\,\sqrt{(1-n_{d})|a_{cd}|^{2}-n_{c}|a_{0}|^{2}}}{|a_{cd}|^{2}-|a_{0}|^{2}}
\end{eqnarray}
 where the first and second line apply if $0<N<1$ or $1<N<2$, respectively.
The energy functional $Q[n_{c},n_{d}]$ is found by minimizing the
above expression with respect to $a_{0},a_{cd}$ and $\varphi$. The
minimum of the functional happens when $a_{cd}=0$ for $0<N<1$, while
for $1<N<2$, it is $a_{0}$ which vanishes. The resulting functional
$Q[n_{c},n_{d}]-\epsilon_{c}\, n_{c}-\epsilon_{d}\, n_{d}\,$ has
the following piece-wise shape:
\begin{eqnarray}
 & - & 2\,|t|\,\sqrt{n_{c}\, n_{d}}\nonumber \\
 & - & 2\,|t|\,\sqrt{(1-n_{c})\,(1-n_{d})}\,+\, U\,(n_{c}+n_{d}-1)
\end{eqnarray}
 where again the first and second line apply if $0<N<1$ or $1<N<2$,
respectively. $Q$ can be easily split into kinetic and interacting
parts, where both must be defined piece-wise. The kinetic term explicitly
shows electron-hole symmetry. The interacting term is non-zero only
if $N>1$, from which a rather simple expression for the exact
$V^{XC}$ can be extracted.


\begin{thebibliography}{References}
\bibitem{Hohenberg64} P. Hohenberg and W. Kohn, 
Phys. Rev. \textbf{136}, B864 (1964).

\bibitem{Kohn65} W. Kohn and L. J. Sham, 
Phys. Rev. \textbf{140}, A1133 (1965).

\bibitem{Perdew81} J. P. Perdew and A. Zunger, 
Phys. Rev. B \textbf{23}, 5048 (1981).

\bibitem{Perdew96} J. P. Perdew, K. Burke, and M. Ernzerhof, 
Phys. Rev. Lett. \textbf{77}, 3865 (1996).

\bibitem{Becke93} A. D. Becke, 
J. Chem. Phys. \textbf{98}, 1372 (1993).

\bibitem{Cohen08b} A. J. Cohen, P. Mori-S�nchez and W. Yang, 
Science \textbf{321}, 792 (2008).

\bibitem{Mori09} P. Mori-S\'anchez, A. J. Cohen and W. Yang, 
Phys. Rev. Lett. \textbf{102}, 066403 (2009).

\bibitem{Schonhammer95} K. Schonhammer, O. Gunnarsson and R. M. Noack,
Phys. Rev. B \textbf{52}, 2504 (1995).

\bibitem{Lieb68} E. H. Lieb and F. Y. Wu, 
Phys. Rev. Lett. \textbf{20}, 1445 (1968).

\bibitem{Yang66} C. N. Yang and C. P. Yang, 
Phys. Rev. B \textbf{150}, 321 (1966).

\bibitem{Lima03} N. A. Lima, M. F. Silva, L. N. Oliveira and K. Capelle,
Phys. Rev. Lett. \textbf{90}, 146402 (2003).

\bibitem{Schenk08} S. Schenk, M. Dzierzawa, P. Schwab and U. Eckern,
Phys. Rev. B \textbf{78}, 165102 (2008).

\bibitem{Verdozzi08} C. Verdozzi, 
Phys. Rev. Lett. \textbf{101}, 166401 (2008).

\bibitem{Kurth10} S. Kurth, G. Stefanucci, E. Khosravi, C. Verdozzi
and E. K. U. Gross, 
Phys. Rev. Lett. \textbf{104}, 236801 (2010).

\bibitem{Burke11a} J. P. Bergfield, Z. Liu, K. Burke and C. Stafford,
arXiv:1106.3104v1

\bibitem{Tsvelick83} A. M. Tsvelick, P. B. Wiegmann, Adv. Phys. \textbf{32},
453 (1983).

\bibitem{Janak78} J. F. Janak, 
Phys. Rev. B \textbf{18}, 7165 (1978).

\bibitem{Perdew83} J. P. Perdew and M. Levy, 
Phys. Rev. Lett. \textbf{51}, 1884 (1983).

\bibitem{Sham83} L. J. Sham and M. Schluter, 
Phys. Rev. Lett. \textbf{51}, 1888 (1983).

\bibitem{Romaniello09} P. Romaniello, S. Guyot and L. Reining, 
J. Chem. Phys. \textbf{131}, 154111 (2009).

\bibitem{Explanation}
Notice that the added quasi-electron will have two
excitation energies. The first will appear if it falls in the same atom where
the first electron was placed since it will feel the mutual Coulomb interaction.
The second excitation energy will appear if it falls in the empty atom, because
then it will feel no Coulomb interaction at all. In contrast, mean field
theory only gives a single excitation energy because the added electron will
interact with half an electron regardless of the atom where it falls.

\bibitem{Rieger95} M. M. Rieger and P. Vogl, 
Phys. Rev. B \textbf{52}, 16567 (1995).

\bibitem{Filippetti03} A. Filippetti and N. A. Spaldin, 
Phys. Rev. B \textbf{67}, 125109 (2003).

\bibitem{Toher09} C. Toher, A. Filippetti, S. Sanvito, and K. Burke,
Phys. Rev. Lett. 95, 146402 (2005)

\bibitem{Hedin65} L. Hedin, 
Phys. Rev. \textbf{139}, A796 (1965).

\bibitem{Aulbur99} W. G. Aulbur, L. Johnson and J. W. Wilkins, Solid
State Physics \textbf{54}, 1 (1999).

\bibitem{Onida02} G. Onida, L. Reining and A. Rubio, Rev. Mod. Phys.
\textbf{74}, 601 (2002).

\bibitem{Bruneval05} F. Bruneval, F. Sottile, V. Olevano, R. Del
Sole and L. Reining, 
Phys. Rev. Lett. \textbf{94}, 186402 (2005).

\bibitem{Wang08} X. Wang, C. D. Spataru, M. S. Hybertsen and A. J.
Millis, Phys. Rev. B \textbf{77}, 045119 (2008).

\bibitem{Thygesen08} K. S. Thygesen and A. Rubio, Phys. Rev. B \textbf{77},
115333 (2008).

\bibitem{HewsonBook} A. C. Hewson, The Kondo Problem to Heavy Fermions
(Cambridge University Press, Cambridge, 1992).

\bibitem{Ferrer87} J. Ferrer, A. Martin-Rodero and F. Flores, Phys.
Rev. B \textbf{36}, 6149 (1987).

\bibitem{Sun04} P. S. Sun and G. Kotliar, 
Phys. Rev. Lett. \textbf{92}, 196402 (2004).

\bibitem{Kotliar06} G. Kotliar, S. Y. Savrasov, K. Haule, V. S. Oudovenko,
O. Parcollet, C. A. Marianetti, 
Rev. Mod. Phsy. \textbf{78}, 866 (2006).

\bibitem{Jacob10} D. Jacob, K. Haule and G. Kotliar, 
Phys. Rev. B \textbf{82}, 195115 (2010).

\bibitem{Korytar11} R. Korytar and N. Lorente, 
J. Phys.: Condens. Matter \textbf{23}, 355009 (2011).

\bibitem{Burke11b} E.M. Stoudenmire, L. O. Wagner, S. R. White and K. Burke, arXiv:1107.2394v1

\bibitem{Levy79} M. Levy, 
Proc. Natl. Acad. Sci. U.S.A. \textbf{76}, 6062 (1979).

\bibitem{Provided}
The minimization procedure described in the main text could be applied even in the hypothetical case where there
would exist several degenerate states within a given box, by just choosing one of these as a representative
of the box.

\bibitem{MahanBook} G. D. Mahan, Many Particle Physics (Plenum Publishing
Corporation, 1981).

\bibitem{Lunquist} L. P. Kadanoff and G. Baym, Quantum Statistical
Mechanics (Addison-Wesley Publishing Company, 1962)

\bibitem{Gritsenko96}
O. V. Gritsenko and E. J. Baerends,
Phys. Rev. A {\bf 54}, 1957 (1996).

\bibitem{Helbig09}
N. Helbig, I. V. Tokatly and A. Rubio,
J. Chem. Phys. {\bf 131}, 224105 (2009).

\bibitem{Perdew82} J. P. Perdew, R. G. Parr, M. Levy and J. L. Balduz,
Phys. Rev. Lett. \textbf{49}, 1691 (1982).

\bibitem{Hewson93} A. C. Hewson, 
Phys. Rev. Lett. \textbf{70}, 4007 (1993). \end{thebibliography}
\end{document}